# Emergence of Diverse Topological States in Ge Doped $MnBi_2Te_4$

Zhijian Shi[1], Shengjie Xu[1], Jianfeng Wang[1,*], Yi Du[1,2,*], Weichang Hao[1,2,*]

[1] *School of Physics, Beihang University, Haidian District, Beijing 100191, China*
[2] *Analysis & Testing Center of Beihang University. Beihang University, Beijing 100191, China*

*E-mails: wangjf06@buaa.edu.cn; yi_du@buaa.edu.cn; whao@buaa.edu.cn

As an ideal platform for studying interplays between symmetry, topology and magnetism, the magnetic topological insulator (MTI) $MnBi_2Te_4$ has attracted extensive attentions. However, its strong *n*-type intrinsic defects hinder the realizations of exotic phenomena. Stimulated by recent discoveries that Ge doping can efficiently tune the position of Fermi level, here we systematically investigate the band evolution and topological phase diagram with doping concentration from MTI $MnBi_2Te_4$ to strong topological insulator $GeBi_2Te_4$. Different from magnetically doped $Bi_2Se_3$, the topology here is determined by competition of two band inversions arising from band folding of two time-reversal invariant momenta between antiferromagnetic and nonmagnetic/ferromagnetic unit cells. By employing a band momentum mapping method, besides the known MTI phase, remarkably, we find two classes of magnetic Dirac semimetal phases at antiferromagnetic state, two classes of Weyl semimetal phases at ferromagnetic state, and an intermediate trivial state at different doping regions. Interestingly, the trivial state can be tuned into a Weyl phase with two coexisting band inversions and extraordinarily long Fermi arcs by a small strain. Our work reveals diverse topological states with intrinsic quantum phenomena can be achieved with great potential for designing future electronic devices.

## I. INTRODUCTION

The past two decades have witnessed tremendous progress in the field of symmetry-protected topological states and materials [1-11], which revolutionizes our understandings in solids. The topological insulator (TI) [1,2,12-17], one of the landmark achievements, characterized by global topological invariants instead of local order parameter, exhibits insulating bulk but conducting boundary states that are protected by time-reversal symmetry (TRS) and immune to backscattering. The extension of topological concepts to semimetals and superconductors enables the realization of particles predicted in high-energy physics, such as Dirac, Weyl, and Majorana fermions, unveiling unique phenomena like Fermi arcs, chiral anomalies and non-Abelian statistics [2-6,18-22]. More recently, the interplay between magnetism and topology has given rise to even more exotic states, including the quantum anomalous Hall effect (QAHE) and axion insulator states [23-27]. These novel topological states are not only of fundamental interest but also hold promise for applications in quantum computing, low-power electronics, spintronics, and other next-generation technologies. Currently, two critical challenges remain in the field of topological matters: first, the search for ideal materials to realize topological states under less demanding conditions, such as replacing magnetically doped TIs with intrinsic magnetic TIs to elevate the temperature for achieving QAHE; and second, the development of controllable phase transitions among multiple topological states, which is crucial for advancing practical applications.

As the first achievable intrinsic magnetic TI (MTI), MnBi$_2$Te$_4$ (MBT) has recently attracted great research interests, which was predicted as an antiferromagnetic (AFM) MTI with a Dirac surface magnetic gap [28-30]. Remarkably, the QAHE [31], topological axion states [32] and topological superconducting states [33] have been observed in its thin-layer samples. However, numerous challenges and controversies exist for experiments in MBT. Many angle-resolved photoemission spectroscopy (ARPES) measurements observed gapless surface states contradicting theoretical predictions [34-36]. More importantly, the experimental reproduction of QAHE in MBT is extremely difficult [31]. These controversies can be related to the existence of a large number of intrinsic Bi/Mn antisite defects [37-39], which also leads to a strong $n$-type doping with Fermi level deep in conduction band for almost all synthetic MBT samples [34-36]. Besides hindering the realization of intrinsic exotic phenomena, meanwhile achieving the transition between multiple magnetic topological states is also a challenge. Very recently, it is found that Ge doping in MBT can efficiently modulate the intrinsic defects and the position of Fermi level, thereby suppressing the $n$-type doping [40,41]. However, the topology of Mn$_{(1-x)}$Ge$_x$Bi$_2$Te$_4$ (MGBT) is unclear. It notes that having the same crystal structure, MBT and GeBi$_2$Te$_4$ (GBT) represent two distinct classes of TIs with completely different band inversions and topological invariants [28, 42]. Doping Ge in MBT may result in the competition of multiple band inversions with emerging diverse topological states when combined with magnetism in MGBT, which is very different from magnetically doped Bi$_2$Se$_3$ [24]. Despite some recent ARPES measurements [40,41,43], the unambiguous band and topological evolutions of MGBT with doping remain unrevealed, that may present an intriguing avenue for further applications with intrinsic quantum phenomena.

In this work, we elucidate the evolution of energy bands and topological phase transitions (TPTs) in MGBT with doping by employing the first-principles calculations. Considering the competition of different band inversions arising from band folding of two time-reversal invariant momenta (TRIM) between AFM and nonmagnetic (NM)/ferromagnetic (FM) unit cells, here we adopt a *band momentum mapping* method, as will be illustrated later. Remarkably, we identify the occurrence of six successive topological phases as doping in the AFM state: MTI, class-I magnetic Dirac semimetal (MDSM), normal insulator (NI), class-II MDSM, MTI, and strong TI (for NM GBT). While for the FM state, which has a slightly higher energy for each doping level, MGBT exhibits corresponding consecutive topological phases with similar transition points as AFM state: class-I Weyl semimetal (WSM) with two Weyl nodes, NI, class-II WSM with eight Weyl nodes, TRS-breaking TI, and strong TI (for GBT). Intriguingly, under a small strain, the NI state can be tuned into a WSM with two band inversions and extraordinarily long Fermi arcs, whose AFM state may produce an exotic magnetic higher-order TI. Our work not only uncovers a rich variety of topological states beyond MBT and GBT, but also provides a platform for achieving multiple TPTs with intrinsic quantum phenomena towards future electronic device designing.

## II. RESULTS & DISCUSSIONS

### A. Band momentum mapping (BMM) and evolution of topology for AFM MGBT.

As revealed by experiments [40,41,43], MGBT has the same tetradymite-type structure as MBT [44] and GBT [45] with space group of $R\bar{3}m$, which is composed of Te-Bi-Te-$M$-Te-Bi-Te

($M$ = Mn/Ge) septuple layers (SLs) van der Waals stacking. Moreover, MGBT exhibits an $A$-type AFM ground state like MBT even at high doping ratios (e.g., $x$ = 0.72 [40] and $x$ = 0.78 [41]), where the $M$ atoms in a SL have the parallel magnetic moments forming a FM order with out-of-plane easy axis, while the adjacent SLs have the opposite magnetization, as shown in Fig. 1(a). Therefore, the AFM MGBT has a 2-SL unit cell and FM/NM MGBT has a 1-SL unit cell [red and black dashed lines in Fig. 1(a) respectively]. As shown in Fig. 1(b), a folding of Brillouin zone (BZ) occurs between these two unit cells. Specifically, the $Z_0$ point in FM/NM BZ ($BZ_0$) is folded to the $\Gamma$ point in AFM BZ, which leads to the mixing of band information around $\Gamma_0$ and $Z_0$.

Let's first illustrate the distinct topological classes of MBT and GBT. In the AFM state of MBT, the TRS is absent, but a composite symmetry combined with time-reversal and fractional translation operations, $S = \tau_{1/2}T$, exists, where $\tau_{1/2}$ represents the half translation operator connecting the two opposite magnetic atoms, as illustrated in Fig. 1(a). The $S$ operator is antiunitary on $k_c$ = 0 plane, similar to $T$ in TIs [27], and also leads to a topological invariant with $Z_2$ = (1, 000) due to a pair of bands inversion with opposite parity induced by spin-orbit coupling (SOC) at the $\Gamma$ point [see Fig. S3(a) [46]]. Thus, the AFM MBT is an MTI. In addition, as predicted by theoretical calculations, the FM state of MBT is a WSM with a pair of Weyl points located along $\Gamma_0Z_0$ [Fig. S3(c) [46]]. While the NM GBT is a TRS-protected strong TI, which has a band inversion occurring at the $Z_0$ point in $BZ_0$ with a topological invariant $Z_2$ = (1, 001) [Fig. S3(b) [46]]. Now let's turn to the topology of MGBT. Considering the 2-SL unit cell for AFM MGBT, it seems that the only band inversion occurs at the same $\Gamma$ point due to the BZ folding of GBT. However, it is not a simple TI model with four bands. Actually, it involves eight bands (including the spin degeneracy). From our calculations (Fig. S3 [46]), the related bands for AFM MBT and 2-SL GBT are compared and schematically plotted in Fig. 1(c). It can be seen that two different groups of band inversions occur for these two systems, which results in completely different parity orderings. Consequently, as the doping ratio $x$ gradually changes, the band inversion within the AFM MGBT will switch, with multiple band inversions/crossings possibly emerging, which, however, is hard to identify.

To characterize the complicated topological electronic structure evolution of the AFM MGBT family, here we develop a *band momentum mapping* (BMM) method, i.e., to map the band in the AFM state to the momentum of NM/FM BZ. It notes that the folding of the bands may modify the dispersion of the electronic structure within reciprocal space without, however, changing the basic characteristics. Therefore, utilizing the orbital distributions of the energy band and the symmetry irreducible representations, such as the eigenvalues of space inversion symmetry (i.e., parity), we can establish the mapping relationship between the energy bands at the $\Gamma$ point in BZ and the momentum at $\Gamma_0/Z_0$ in $BZ_0$. It also notes that the BMM method is very different from the band unfolding method, where the latter unfolds the supercell bands back to the primitive cell's BZ, while the BMM maps the origin of the primitive cell's momentum within the supercell bands. Thus, the BMM method can be regarded as complementary to the band unfolding method, and it excels at preserving all information of the energy bands, such as parity and other properties. By applying this method, we can independently track the evolution of the energy band around $\Gamma$ from $\Gamma_0/Z_0$ with the doping for all AFM MGBT (more details can refer to the Supplementary Note 2 in SM [46]).

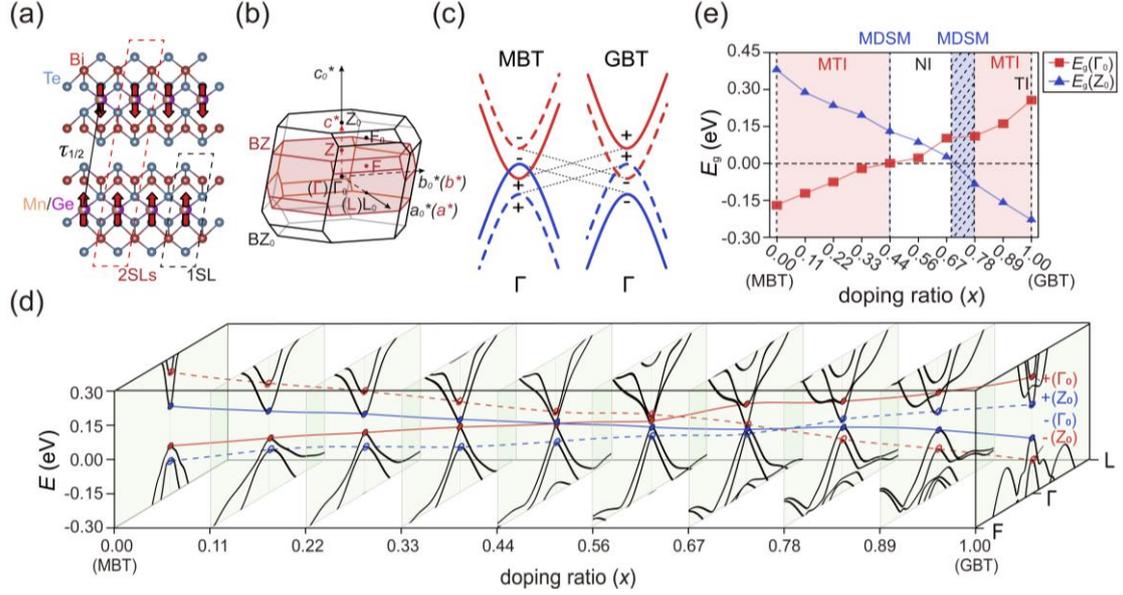

FIG 1. The evolution of band topology in AFM MGBT. (a) Crystal structure of AFM MGBT, where Bi, Te and Mn/Ge atoms are represented by different colored spheres. The AFM order of adjacent Mn/Ge layers is marked by red arrows, which can be related by a fractional translation operation $\tau_{1/2}$. (b) BZ folding between AFM unit cell [red dashed lines in (a)] and FM/NM unit cell [black dashed lines in (a)], whose BZs are plotted by red (BZ) and black lines ($BZ_0$) respectively. Specifically, the $Z_0$ point in $BZ_0$ is folded to $\Gamma$ in BZ. (c) Schematic comparison of band inversions between AFM MBT and 2-SL GBT. The parities at $\Gamma$ are labelled. (d) Energy band evolution of AFM MGBT with momentum $\Gamma_0/Z_0$ of $BZ_0$ mapping as a function of doping ratio $x$, where the band parities at $\Gamma$ are marked and the solid/hollow circles respectively represent the states from $\Gamma_0/Z_0$ of $BZ_0$. The red/blue colors represent the Bi/Te orbital contributions and the solid/dash lines mark the bands evolution from $\Gamma_0/Z_0$ of $BZ_0$ respectively. (e) Topological phase diagram of AFM MGBT with energy gap evolution of $\Gamma_0$ and $Z_0$ as a function of doping ratio $x$.

In our first-principles calculations, all doping structures are created through building various $3 \times 3 \times 2$ supercells (see Methods section and Supplementary Note 1 in SM [46]), where the AFM $S$, intralayer space inversion ($P_1$) and combined interlayer inversion and time-reversal ($P_2T$) symmetries are preserved, which are essential for some topological definitions. The calculated band structures with ten different doping levels from MBT ($x = 0$) to GBT ($x = 1$) are shown in Fig. 1(d). By applying the strategy of BMM method, the band states originating from $\Gamma_0$ (solid circles) and $Z_0$ (hollow circles) around the $\Gamma$ point are marked, where the red and blue colors indicate the Bi-$p$ and Te-$p$ orbital contributions respectively, that are necessary for building the BMM relationship. From the red/blue solid/dashed lines in Fig. 1(d), the band evolutions from $\Gamma_0$ and $Z_0$ are opposite. Figure 1(e) illustrates the bandgap changes from $\Gamma_0$ and $Z_0$, which reveals their opposite evolution trends.

From Figs. 1(d) and 1(e), the topological phases and TPTs for the AFM MGBT with doping can be clearly elucidated. As $x$ increases, the band edges from $\Gamma_0$ approach each other, undergo inversion, and subsequently deviate from each other, resulting in a transition of the energy gap of $\Gamma_0$ [$E_g(\Gamma_0)$] from a topologically nontrivial to a trivial region. An opposite process occurs for the bandgap of $Z_0$ [$E_g(Z_0)$]. Specifically, during the region from $x = 0$ to $x = 0.44$, the odd-parity Te band inverts with even-parity Bi band from $\Gamma_0$, aligning with the AFM MBT with the topological invariant $Z_2 = (1, 000)$. Around $x = 0.44$, these two parity-opposite bands touch and merge at $\Gamma$ to form a critical crossing point, i.e., class-I MDSM state, which is a TPT point from a MTI state to a NI state [the red and blue solid lines cross at this point in Fig. 1(d)]. As the doping ratio increases from $x = 0.44$ to 0.67, the $E_g(\Gamma_0)$ continues to increase and the $E_g(Z_0)$ decreases, with no band inversion occurring at both the $\Gamma_0$ and $Z_0$ points. In the region from $x = 0.67$ to 0.78, the odd-parity band from $Z_0$ passes through the even-parity band from $Z_0$ and odd-parity band from $\Gamma_0$ [the red dashed line passes through the blue dashed and solid lines at this region in Fig. 1(d)], bringing about a band inversion at $Z_0$. Remarkably, the band crossing along the $\Gamma Z$ path results in the formation of a quadruple degenerate Dirac point, thereby establishing a class-II MDSM state. After this region, the AFM order MGBT remains in the MTI state until the magnetic elements vanish, restoring TRS to achieve a strong TI state from $x = 0.78$ onwards to $x = 1$.

Consequently, six successive topological states emerge from $x = 0$ (MBT) to $x = 1$ (GBT) for AFM MGBT: MTI, class-I MDSM, NI, class-II MDSM, MTI, and TI. It is noteworthy that the specific TPT points may be related to the constructed supercell structure, but the overall band evolution and phase diagram should remain essentially unchanged. In the next two sections, we will provide a comprehensive elucidation of the topological properties associated with these distinct insulating and semimetallic phases.

### B. Topologically distinct insulator states in MGBT.

Besides the MTI MBT and TI GBT, for MGBT family system, the insulating phase predominantly prevails. Here we choose three representatives for insulator state regions, specifically at $x = 0.22$, 0.56 and 0.89, to demonstrate their topological classifications and properties.

The first phase, with the doping ratio extending from $x = 0$ to 0.44, is topologically homeomorphic to MBT because of their similar band structures. As depicted in Fig. 2(a) and Fig. S4, the parity distribution of band edges for MGBT at $x = 0.22$ aligns precisely with that of MBT. Consequently, with the $S$ symmetry, MGBT at $x = 0.22$ exhibits a $Z_2 = 1$ invariant based on the evolution of wannier charge centers (WCCs), indicative of an MTI state [see Fig. S5(a) [46]]. The calculated surface states for AFM MGBT at $x = 0.22$ are presented in Figs. 2(d) and 2(g). On the (001) surface, the broken $S$ symmetry produces a Dirac-type topological surface state (TSS) with a magnetic gap. While, the preserved $S$ symmetry on the $(1\bar{1}0)$ surface [also on the (110) surface] protects a gapless Dirac cone of the TSS. Representing the second NI phase, the AFM MGBT at $x = 0.56$ is characterized as a topological trivial state, lacking any $Z_2$ topological charge or topological TSS, as illustrated in Figs. 2(b), 2(e), 2(h) and Figs. S4(e), S5(b), S5(e) [46].

In the third MTI phase for MGBT at $x = 0.89$, as illustrated in Fig. 2(c), the negative parity band from $Z_0$ inverts with positive parity band from $Z_0$ around the $\Gamma$ point. Despite having the different orders of band parities, the topological indices for both $x = 0.89$ and $x = 0.22$ are identical, and they possess topologically equivalent surface states and evolutions of WCC, as shown in Figs. 2(f), 2(i) and Fig. S5(c) [46]. However, the multitudinous substitution of magnetic Mn atoms with non-magnetic Ge atoms has led to a significant weakening of the magnetism, resulting in a reduction of the magnetic gap (~0.8 meV) on the (001) surface.

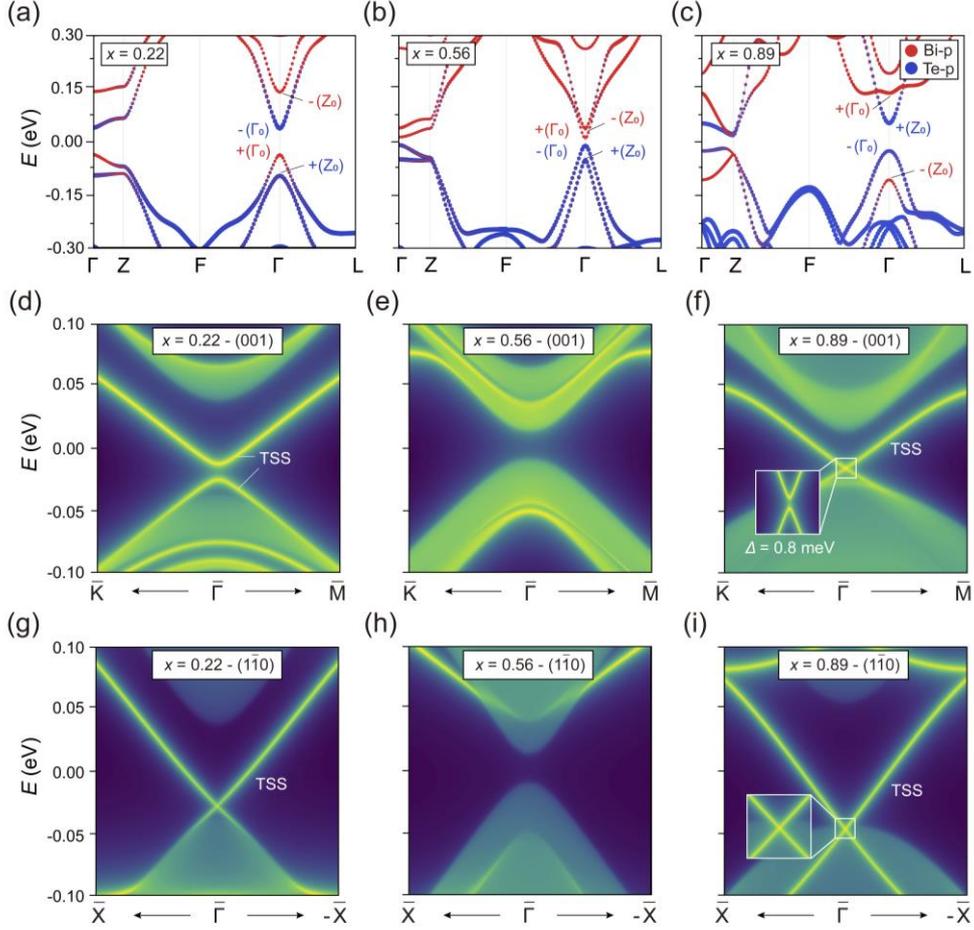

FIG 2. MTI and NI states in AFM MGBT with $x$ = 0.22, 0.56, 0.89. (a-c) Bulk energy bands with Bi/Te $p$ orbitals projections. The parities and mapped momenta ($\Gamma_0/Z_0$) at the $\Gamma$ point are marked. (001) surface bands (d-f) and ($1\bar{1}0$) surface bands (g-i). The topological surface states (TSSs) are labelled. Insets in (f) and (i) are enlarged bands, revealing a gapped and gapless Dirac cone respectively.

### C. Two classes of MDSM states in MGBT.

The Dirac semimetals (DSMs) have been widely studied in NM materials that possess both space inversion $P$ and time-reversal $T$ symmetries [51]. The quadruply degenerate Dirac point (DP) usually requires protection from additional crystalline symmetries, which has been demonstrated in systems at the critical point of TPT, with symmetry-enforced degeneracy at TRIM, such as $BiO_2$ [52], and along high-symmetry line with rotation symmetry, such as $Na_3Bi$ and $Cd_3As_2$ [19,50]. While the DSMs are rarely observed in magnetic systems due to the breaking of TRS, the recent studies found that the DP can exist in AFM system with combined $PT$ symmetry [50], named MDSM. Here we show two classes of MDSM in MGBT at $x$ = 0.44 and $x$ = 0.67-0.78, respectively.

The class-I MDSM for MGBT at $x$ = 0.44 belongs to the critical point of TPT. As shown in Fig. 3(a), two energy bands with opposite parities from $\Gamma_0$ touch and become a DP at the $\Gamma$ point. As indicated in Figs. 1(d) and 1(e), the class-I MDSM acts as a TPT boundary from MTI phase to NI phase.

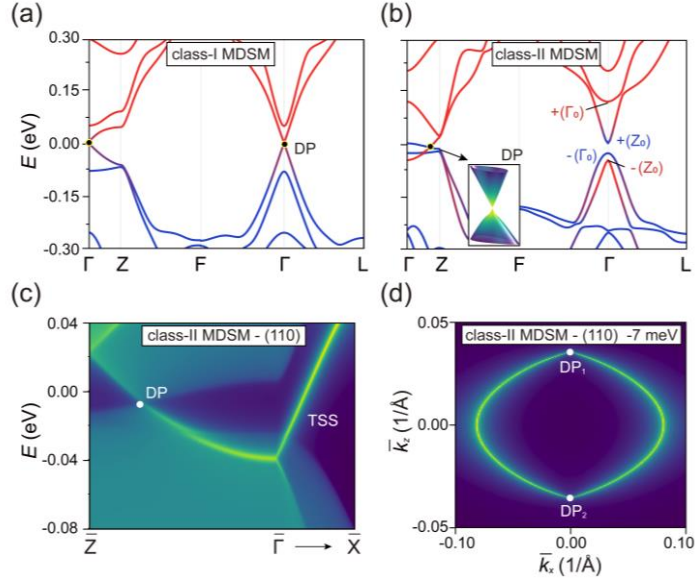

FIG 3. Two classes of MDSM states in MGBT. (a, b) Bulk band structures of class-I and class-II MDSM for MGBT with $x$ = 0.44 and $x$ = 0.67-0.78, respectively. The Dirac point (DP) and mapped momenta at Γ are labelled. Inset of (b) plots the two-dimensional bands around DP on the $k_x$-$k_y$ plane. (c) Surface band and (d) iso-energy surface on the (110) surface of class-II MDSM.

Within the range of $x$ = 0.67-0.78, the AFM MGBT is named class-II MDSM, which has two DPs along high-symmetry line, similar to $Na_3Bi$. A representative band is shown in Fig. 3(b), where the supercell of $x$ = 0.78 under a small strain (~0.37%) is employed (see Supplementary Note 4 [46]). The band primarily composed of Bi-$p$ orbitals (red color) intersects with two bands that are mainly composed of Te-$p$ orbitals (blue color), creating two nodes located on the ΓZ path close to the Fermi level. Furthermore, the node near the Fermi level [marked with a dot in Fig. 3(b)], exhibits Dirac linear dispersion in every direction within the $k_x$-$k_y$ plane, as illustrated in the inset of Fig. 3(b). The inversion between bands of opposite parity from $Z_0$, evidenced by the momentum mapping in Figs. 3(b) and 1(d), along with the evolution of WCC depicted in Fig. S6(c) [46], confirms that this MDSM system carries a $Z_2$ topological charge. As demonstrated in Fig. 3(c), a TSS is observed connecting the DP along the $\overline{ΓZ}$ path and has a linear dispersion along the $\overline{ΓX}$ path on the (110) surface. Consequently, its contour surface at the Fermi level features a pair of Fermi arcs connecting the two DPs, just like $Na_3Bi$.

Let's turn to the symmetry protection of class-II MDSM in MGBT. The AFM state of MGBT lacks $T$ but possesses a combined $P_2T$ symmetry that maintains global double energy band degeneracy [53]. Also similar to the NM $Na_3Bi$, the DPs along ΓZ are protected by the rotation symmetry. This is confirmed by the eigenvalue calculations of $C_3$ rotation in a strained parent MBT with the unit cell [see Figs. S6(a), S7(a) [46]], due to the missing of $C_3$ rotational symmetry for the built supercell of MGBT, which also results in a tiny gap of Dirac node in MGBT. It shows that the two crossing energy bands forming the DP exhibit distinct eigenvalues of $C_3$ rotation [Fig. S6(a) [46]]. Considering the preservation of all the crystal symmetries of parent phase in the random doped alloy, it is noted that such an equivalent investigation method by using strained parent unit cell to replace the doped supercell is very useful, when exploring some properties related to symmetries that are lost for supercell. This equivalent investigation can be implemented by establishing a correlation of energy bands between the doped supercell and parent unit cell through employing our BMM approach, and it has been proven to be effective in MGBT, as most topological phases can be replicated using strained MBT (see Supplementary Note 4 [46]).

### D. Topological phase diagram for FM MGBT.

Based on our DFT calculations, all the MGBT family materials have AFM ground states, but only with a slightly higher energy for the FM states [see Fig. S2(b) [46]]. Here we demonstrate the possible topological states for FM MGBT with different doping levels. Due to the consistent BZ between FM and NM states, the evolution of energy bands and topology under different doping levels can be directly obtained through orbital projection without the need for BMM approach (see Fig. S9 for projected bands [46]). The bandgap evolutions of $\Gamma_0$ and $Z_0$ are shown in Fig. 4(a), which exhibits five consecutive topological phases with similar transition points as AFM state: class-I WSM with two Weyl nodes, NI, class-II WSM with eight Weyl nodes, TRS-breaking TI (TBTI), and strong TI (for GBT), as roughly illustrated below. At the region of $x$ = 0-0.44, two bands from Bi and Te orbitals are inverted across each other to form a pair of Weyl points (WPs) located at two sides of $\Gamma_0$ along the $\Gamma_0Z_0$ path. As the doping ratio increases from $x$=0.44 to 0.67, there is no band inversion, and the FM MGBT is a NI. In the region of $x$ = 0.67-0.78, the DPs of the AFM state split into WPs in its FM state, thereby producing a class-II WSM with a total of eight WPs. Within the region between $x$ = 0.78 to $x$ = 1, the spin splitting of energy bands is small and both spin bands undergo inversion around $Z_0$, forming a TI phase with broken TRS, named TBTI, until the magnetism completely disappears, resulting in a strong TI phase for GBT.

Three representative bands for class-I WSM, class-II WSM and TBTI phases are shown in Figs. 4(b)-4(d) respectively. The class-I WSM ($x$ = 0.22) has a similar band as parent FM MBT [Fig. 4(b)], wherein a pair of spin-opposite bands cross around the $\Gamma_0$ point creating two WPs. Its nonzero Chern number ($C$ = 1) on $k_c$ = 0 plane and TSS on the (110) surface are depicted in Figs. S10(a) and S11(a)-S11(c) [46]. We focus on class-II WSM now. When transitioning from an AFM state to a FM state in the region of $x$ = 0.67-0.78, the DPs of class-II MDSM will split into a series of WPs. Here an equivalent investigation on strained parent MBT is implemented considering the necessary $C_3$ rotation symmetry [Fig. 4(c)]. Unlike the common scenario where a DP splits into a pair of WPs with opposite chirality, here two DPs evolves into *eight* WPs: Two WPs are distributed along the $\Gamma_0Z_0$ path, while the other six exhibit triple rotation with each inversion partner having opposite chirality, as illustrated in Fig. 4(e). Interestingly, such class-II WSM has a high Chern number of $C$ = 2 on the $k_c$ = 0 plane [Fig. S12(b) [46]], which are consistent with the two branches of chiral TSS on its (100) surface and multiple Fermi arcs connecting WPs as shown in Figs. 4(f) and 4(g). Thus, this system could be potentially turned into a Chern insulator with a high Chern number, if the degeneracy of WPs is disrupted by additional symmetry-broken perturbations.

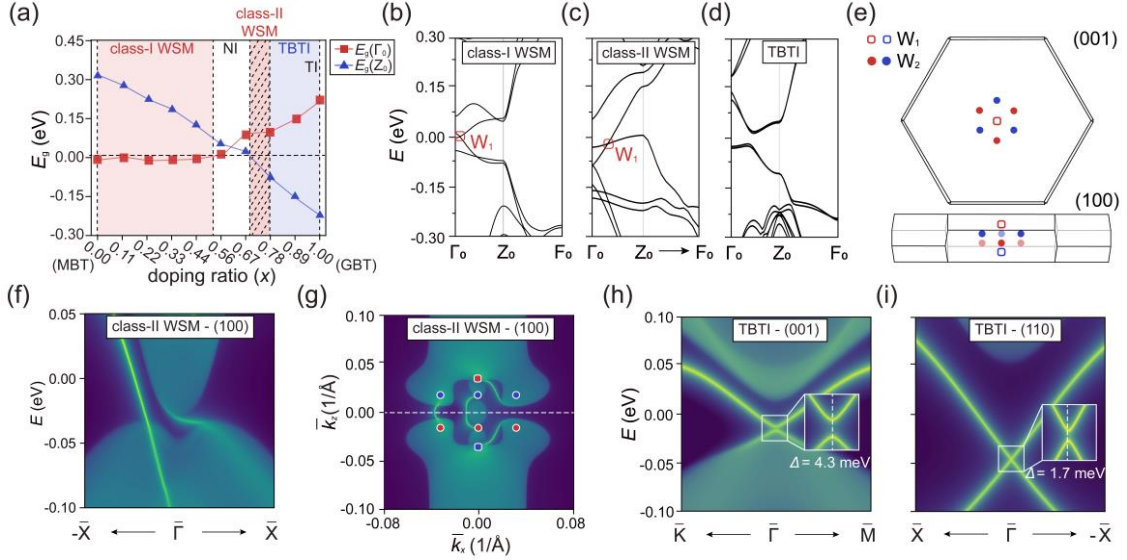

FIG 4. Multiple topological states in FM MGBT. (a) Topological phase diagram for FM MGBT. (b-d) Band structures of MGBT with class-I WSM, class-II WSM and TRS-breaking TI (TBTI) states at three different doping level. (e) Distributions of Weyl points (WPs) in BZ, where the in-plane distances of all WPs relative to the center of BZ have been proportionally enlarged. WPs further forward in the (100) direction are shown in darker colors, while those further back are in lighter colors. (f, g) Surface band and Fermi arcs of class-II WSM state on the (100) surface. (h, i) (001) and (110) surface bands for TBTI MGBT. The insets are enlargements around Dirac point, which indicate a small gap.

For $x = 0.89$, the weakening of magnetism results in a distinct manifestation of the FM state in MGBT compared to $x = 0.22$: the band spin splitting is significantly reduced due to the overall decrease in magnetism, leading to simultaneous band inversion at the $Z_0$ point for both spins. As illustrated in Figs. 4(h) and 4(i), the FM MGBT with $x = 0.89$ exhibits a Dirac-type surface state on both the (001) and (110) surfaces, akin to that observed in TI. Additionally, the WCC evolution on $k_c = \pi$ plane displayed a $Z_2 = 1$ topological invariant [Fig. S10(f) [46]]. But, the introduction of magnetism disrupts the TRS, leading to the absence of double degeneracy (i.e., forming a tiny gap) for both TSS and WCC at the TRIM points. Consequently, the FM MGBT with $x = 0.89$ is designated as a TBTI, reflecting a unique characteristic within the diagram of topological phases.

**E. WSM with two band inversions and potential magnetic high-order topology.**

From all the results above, the diverse topological states in MGBT originate from the alternating occurrence of band inversions from $\Gamma_0$ and $Z_0$. An intriguing question is whether the two band inversions can coexist in MGBT, which is particularly fascinating since it may involve the interplay between multiple topological phases or the emergence of higher-order topology [54]. Although they do not occur simultaneously in MGBT in all doping region, we notice that within the NI phase region ($x = 0.44$-$0.67$), the trivial energy gaps at $\Gamma_0$ and $Z_0$ of MGBT are both exceptionally small. It means that an appropriate strain might be able to induce simultaneous band inversions at both points, and thus achieving a higher-order topology.

We have tried to implement this idea by applying an in-plane and/or out-of-plane strain to a FM MGBT with $x = 0.56$, which is a NI in the absence of strain. Its topological phase diagram under different strains is shown in Fig. 5(a). Compressive strain will diminish the band gap, instigating band inversions at the $\Gamma_0$ and $Z_0$ points sequentially, which precipitates the emergence of WPs and drives the system back into the WSM phase. Importantly, it is plausible for an appropriately large compressive strain to induce simultaneous band inversions at both points, thereby catalyzing the formation of a novel WSM2 phase with two pairs of WPs [red region in Fig. 5(a)].

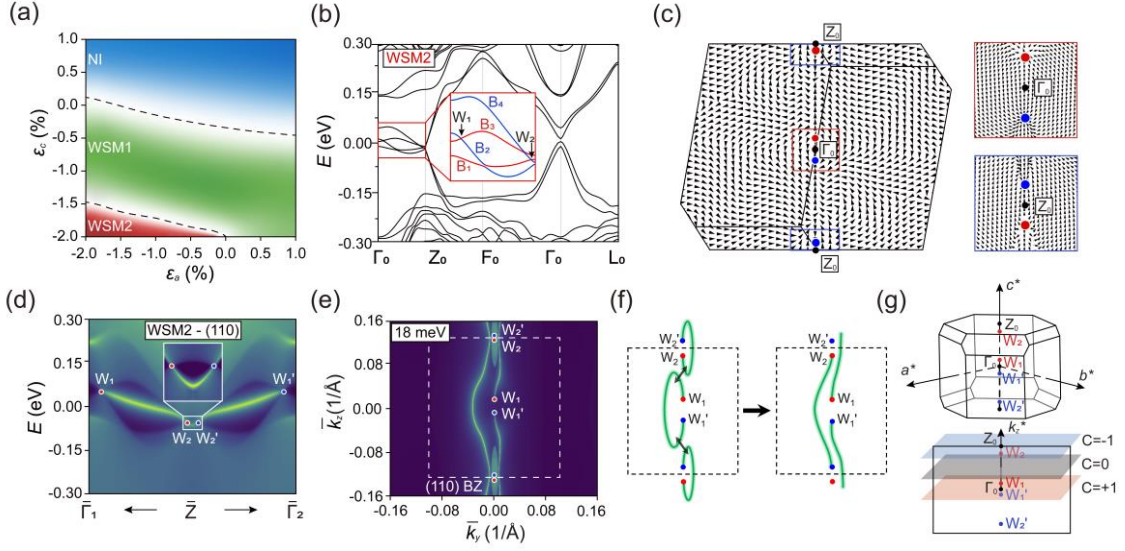

FIG 5. WSM state with two band inversions under strain. (a) Topological phase diagram of FM MGBT with $x = 0.56$ under additional strains, where $\varepsilon_a$ ($\varepsilon_c$) indicates the in-plane (out-of-plane) strain. WSM1 (WSM2) means the WSM phase having one (two) band inversions with a (two) pair of WPs. (b) Bulk band structure of WSM2 phase. In the magnified image, two pairs of spin-opposite bands $B_3/B_2$ and $B_1/B_4$, both marked in red/blue, intersect at two WPs, $W_1$ and $W_2$. (c) Berry curvature and distribution of WPs at the $k_x = 0$ plane in $BZ_0$. (d) Surface band on the (110) surface along the $\bar{\Gamma}\bar{Z}$ path and (e) its iso-energy surface with energy at 18meV. The chirality-opposite WPs are marked by red/blue dots. (f) Schematic plot of coupling between two branches of Fermi arcs connecting $W_1/W_1'$ and $W_2/W_2'$ to form the presented Fermi arc in (e). (g) Calculated Chern numbers on the $k_c = 0, \pi/2, \pi$ planes within the $BZ_0$, which separate $W_1'$, $W_1$, $W_2$, $W_2'$, respectively.

Here, a 2% triaxial compressive strain on the FM MGBT with $x = 0.56$ is selected as a representative for an in-depth exploration of this WSM2 phase. As shown in Fig. 5(b), both band inversions around $\Gamma_0$ and $Z_0$ create two distinct WPs, designated as $W_1$ and $W_2$, situated along the high-symmetry $\Gamma_0 Z_0$ path. It is evident from the inset of Fig. 5(b) that $W_1$ and $W_2$ arise from the different crossings of distinct bands ($B_2/B_3$ for $W_1$ crossing and $B_1/B_4$ for $W_2$ crossing), with no symmetrical relationship existing between these two pairs of WPs. That is to say, $W_1$ and $W_2$ appear to be regarded as two completely independent WPs located at different energy positions. According to the Berry curvature and chirality calculations depicted in Fig. 5(c) and Fig. S14 [46], WPs residing in the same half of the $BZ_0$ share the same chirality.

Figure 5(d) demonstrates the calculated (110) surface band of this WSM2 state. Surprisingly, the TSS does not directly connect $W_1$ and $W_1'$ by passing through the $\bm{k_c} = 0$ plane/line, but appears to connect to $W_2/W_2'$ with the same chirality. As shown in Fig. 5(e), its iso-energy surface also demonstrates that the Fermi arc starts from $W_1$ and terminates at $W_2$ or passes through the boundary of BZ ($k_c = \pi$) to connect $W_1'$, which is very different from FM MBT. At the same time, there is another Fermi arc spanning across the entire BZ. When examined independently, each pair of WPs with opposite chirality is connected by a single TSS or Fermi arc [see inset of Fig. 5(d) and Fig. S13 [46]]. From the evolution of Fermi arc with varying energies, it can be inferred that the Fermi arcs displayed in Fig. 5(e) are the result of the mutual coupling between two branches of Fermi arcs connecting $W_1/W_1'$ and $W_2/W_2'$, respectively, as schematically plotted in Fig. 5(f). Consequently, extraordinarily long Fermi arcs nearly spanning BZ can unexpectedly emerge in the strained MGBT system, which is highly advantageous for experimental observation. Thus, this system also provides a platform for studying the interactions between multiple WPs. The calculated Chern numbers based on WCC evolution (Fig. S15 [46]) on the $k_c = 0, \pi/2, \pi$ planes are $C = +1, 0, -1$ respectively, as depicted in Fig. 5(g). It reveals that the Chern number will change by 2 from $k_c = 0$ plane to $k_c = \pi$ plane, which is consistent with the two band inversions and double WPs with identical chirality along $\Gamma_0 Z_0$.

Finally, when the same strain is applied to the AFM MGBT with $x = 0.56$, the two band inversions from $\Gamma_0$ and $Z_0$ occur simultaneously at the $\Gamma$ point due to the BZ folding. However, according to the parity and WCC calculations (Fig. S16 [46]), the topological property of this system is characterized by a topological invariant $Z_4 = 0$, different from the recently discovered second-order TI with $Z_4 = 2$. Nevertheless, its TSS still exists but exhibit gapped features on both (001) and (110) surfaces (Fig. S17 [46]). Besides the $Z_4$ topological index, some other higher-order topological states and topological invariants have been recently proposed [55], such as the two-dimensional SSH model [56] and multipole moment indicators [57]. The further exploration of AFM higher-order topology deserves future research.

### III. CONCLUSION

Diverse topological states can emerge in Ge doped MBT with different doping ratios, including MTI, strong TI, two classes of MDSM, NI for AFM MGBT, and two classes of WSM, TBTI, NI for FM MGBT. The topological evolution arises from the competition of two band inversions with different momenta origins, as revealed by our DFT calculations and developed BMM method. It is highlighted that the BMM method can be regarded as complementary to the band unfolding approach, and is well-suited for investigating the band evolution between different unit cells. Compared to the existing topological states in the parent MBT and GBT, the unique class-II MDSM and class-II WSM states in MGBT exhibit more intriguing and distinctive properties that deserve further attention for studying the interplay between crystalline symmetry, magnetism and topology. On the experimental front, it has been demonstrated that Ge doping can effectively modulate the Fermi level of MBT [41]. Notably, the MGBT system approaches neutrality around $x = 0.46$, and within this range, the magnetotransport measurements reveal intricate TPT behaviors [40]. Recently, other theoretical calculations have also predicted MDSM and WSM, yet these works neglect the critical role of symmetry in topology and thus cannot reflect topological evolution [41, 43]. Moreover, based on our results, in the vicinity of this doping range, a trivial NI phase of MGBT can be transformed into a novel WSM with two band inversions and interacting Fermi arcs or a higher-order MTI state by applying a small strain. Our work reveals a promising platform for the exploration of a multitude of magnetic topological phenomena and future electronic device applications.

## IV. METHODS

### A. Computational methods

The first-principles calculations were performed by using the Vienna ab initio simulation package (VASP) with the projector augmented wave approach (PAW) and the Perdew-Burke-Ernzerh (PBE) generalized-gradient approximation (GGA) [58-60]. The cutoff energy of the plane-wave expansion of the basic functions was set to 360 eV. 13 × 13 × 3 (13 × 13 × 5) and 5 × 5 × 3 (5 × 5 × 5) Γ-centered Monkhorst-Pack $k$-point meshes were used to sample the BZ for 2-SL (1-SL) MBT/GBT unit cells and 3 × 3 × 2 (3 × 3 × 1) MGBT supercells, respectively. All the crystal structures were fully relaxed with DFT-D3 van der Waals correction [61] until the residual force of each atom is smaller than 0.01 eV/Å. The SOC effect is considered in our electronic structure calculations. The Mn-$d$ orbitals were treated within the DFT+U approach [62] using U = 4.0 eV, which results in a magnetic moment of 5.0 μB on Mn atoms in all self-consistent calculations. The tight-binding Hamiltonians based on the maximally localized Wannier functions (MLWF) [63] with Bi-$p$ and Te-$p$ orbitals are constructed to further calculate various topological properties by using the WannierTools package [64]. The symmetry analysis is performed by employing the WannSymm package [65].

### B. Supercell structures for doping simulation

The doping effect of Ge in MBT is simulated by supercell structure calculations. It is noted that the virtual crystal approximation (VCA) [66] is not applicable to our system, because the VCA method is usually limited to mixing elements within the same group or period and the Mn ($3d^5 4s^2$) and Ge ($4s^2 4p^2$) possess entirely different valence electron configurations. In this work, various 3 × 3 × 2 supercell structures are built, where the AFM $S$, intralayer space inversion ($P_1$) and combined interlayer inversion and time-reversal ($P_2T$) symmetries are preserved. Therefore, the doping ratio is $x = n/9$ with integer $0 \leq n \leq 9$, i.e., $x$ = 0.00, 0.11, 0.22, ..., 0.89 and 1.00. For each doping level, the structure with the lowest energy is chosen for the further electronic structure calculation (see Supplementary Note 1 [46]). Additionally, the uniaxial and triaxial strains are also applied to some doped MGBT systems or parent MBT to analyze the formation of certain topological states (see Supplementary Note 4 [46]).

## ACKNOWLEDGMENTS

This work was supported by the National Natural Science Foundation of China (Grant Nos. 12474217, 52473287, 12004030, 12004021 and 12274016), Beijing Natural Science Foundation (Grant No. 1242023), the Fundamental Research Funds for the Central Universities and the National Key R&D Program of China (2018YFE0202700 and 2022YFB3403400).

**References:**

[1] M. Z. Hasan, and C. L. Kane, *Colloquium: Topological Insulators*, Rev. Mod. Phys. **82**, 3045 (2010).

[2] X.-L. Qi, and S.-C. Zhang, *Topological Insulators and Superconductors*, Rev. Mod. Phys. **83**, 1057 (2011).

[3] C. K. Chiu, J. C. Y. Teo, A. P. Schnyder, and S. Ryu, *Classification of Topological Quantum Matter with Symmetries*, Rev. Mod. Phys. **88**, 035005 (2016).

[4] A. Bansil, H. Lin, and T. Das, *Colloquium: Topological Band Theory*, Rev. Mod. Phys. **88**, 021004 (2016).

[5] N. P. Armitage, E. J. Mele, and A. Vishwanath, *Weyl and Dirac Semimetals in Three-Dimensional Solids*, Rev. Mod. Phys. **90**, 015001 (2018).

[6] B. Q. Lv, T. Qian, and H. Ding, *Experimental Perspective on Three-Dimensional Topological Semimetals*, Rev. Mod. Phys. **93**, 025002 (2021).

[7] A. A. Burkov, *Topological Semimetals*, Nat. Mater. **15**, 1145 (2016).

[8] B. Bradlyn, L. Elcoro, J. Cano, M. G. Vergniory, Z. Wang, C. Felser, M. I. Aroyo, and B. A. Bernevig, *Topological Quantum Chemistry*, Nature **547**, 298 (2017).

[9] T. Zhang, Y. Jiang, Z. Song, H. Huang, Y. He, Z. Fang, H. Weng, and C. Fang, *Catalogue of Topological Electronic Materials*, Nature **566**, 475 (2019).

[10] F. Tang, H. C. Po, A. Vishwanath, and X. Wan, *Comprehensive Search for Topological Materials Using Symmetry Indicators*, Nature **566**, 486 (2019).

[11] M. G. Vergniory, L. Elcoro, C. Felser, N. Regnault, B. A. Bernevig, and Z. Wang, *A Complete Catalogue of High-Quality Topological Materials*, Nature **566**, 480 (2019).

[12] C. L. Kane, and E. J. Mele, *Quantum Spin Hall Effect in Graphene*, Phys. Rev. Lett. **95**, 226801 (2005).

[13] C. L. Kane, and E. J. Mele, *$Z_2$ Topological Order and the Quantum Spin Hall Effect*, Phys. Rev. Lett. **95**, 146802 (2005).

[14] B. A. Bernevig, T. L. Hughes, and S.-C. Zhang, *Quantum Spin Hall Effect and Topological Phase Transition in HgTe Quantum Wells*, Science **314**, 1757 (2006).

[15] C. L. Kane, and E. J. Mele, *Topological Insulators in Three Dimensions*, Phys. Rev. Lett. **98**, 106803 (2007).

[16] H. Zhang, C.-X. Liu, X.-L. Qi, X. Dai, Z. Fang, and S.-C. Zhang, *Topological Insulators in $Bi_2Se_3$, $Bi_2Te_3$ and $Sb_2Te_3$ with a Single Dirac Cone on the Surface*, Nat. Phys. **5**, 438 (2009).

[17] W. A. Benalcazar, B. A. Bernevig, and T. L. Hughes, *Quantized Electric Multipole Insulators*, Science **357**, 61 (2017).

[18] X. Wan, A. M. Turner, A. Vishwanath, and S.Y. Savrasov, *Topological Semimetal and Fermi-Arc Surface States in the Electronic Structure of Pyrochlore Iridates*, Phys. Rev. B **83**, 205101 (2011).

[19] Z. Wang, H. Y. Sun, X.-Q. Chen, C. Franchini, G. Xu, H. Weng, X. Dai, and Z. Fang, *Dirac Semimetal and Topological Phase Transitions in $A_3Bi(A=Na,K,Rb)$*, Phys. Rev. B **85**, 195320 (2012).

[20] H. Weng, C. Fang, Z. Fang, B. A. Bernevig, and X. Dai, *Weyl Semimetal Phase in Noncentrosymmetric Transition-Metal Monophosphides*, Phys. Rev. X **5**, 011029 (2015).

[21] C. W. J. Beenakker, *Search for Majorana Fermions in Superconductors*, Annu. Rev. Condens. Matter Phys. **4**, 113 (2013).

[22] J. Alicea, *New Directions in the Pursuit of Majorana Fermions in Solid State Systems*, Rep. Prog. Phys. **75**, 076501 (2012).

[23] R. Yu, W. Zhang, H.-J. Zhang, X. Dai, and Z. Fang, *Quantized Anomalous Hall Effect in Magnetic Topological Insulators*, Science **329**, 61 (2010).

[24] C.-Z. Chang, J. Zhang, X. Feng, J. Shen, Z. Zhang, M. Guo, K. Li, Y. Ou, P. Wei, L.-L. Wang *et al.*, *Experimental Observation of the Quantum Anomalous Hall Effect in a Magnetic Topological Insulator*, Science **340**, 167 (2013).

[25] R. Li, J. Wang and X.-L. Qi, and S.-C. Zhang, *Dynamical Axion Field in Topological Magnetic Insulators*, Nat. Phys. **6**, 284 (2010).

[26] D. Xiao, J. Jiang, J.-H. Shin, W. Wang, F. Wang, Y.-F. Zhao, C. liu, W. Wu, M. H. Chan N. Samarth, and C.-Z. Chang, *Realization of the Axion Insulator State in Quantum Anomalous Hall Sandwich Heterostructures*, Phys. Rev. Lett. **120**, 056801 (2018).

[27] R. S. K. Mong, A. M. Essin, and J. E. Moore, *Antiferromagnetic Topological Insulators*, Phys. Rev. B **81**, 245209 (2010).

[28] J. Li, Y. Li, and S. Du, Z. Wang, B.-L. Gu, S.-C. Zhang, K.He, W. Duan and Y. Xu, *Intrinsic Magnetic Topological Insulators in van der Waals Layered $MnBi_2Te_4$-Family Materials*, Sci. Adv. **5**, eaaw5685 (2019).

[29] D. Zhang. M. Shi, T. Zhu, D. Xing, H. Zhang, and J. Wang, *Topological Axion States in the Magnetic Insulator $MnBi_2Te_4$ with the Quantized Magnetoelectric Effect*, Phys. Rev. Lett. **122**, 206401 (2019).

[30] M. M. Otrokov, I. I. Klimovskikh and H. Bentmann, D. Estyunin, A. Zeugner, Z. S. Aliev, S. Gaß, A. U. B. Wolter, A.V. Koroleva *et al.*, *Prediction and Observation of an Antiferromagnetic Topological Insulator*, Nature **576**, 416 (2019).

[31] Y. Deng, Y. Yu and M. Z. Shi, Z. Guo, Z. Xu, J. Wang, X. H. Chen, and Y. Zhang, *Quantum Anomalous Hall Effect in Intrinsic Magnetic Topological Insulator* $MnBi_2Te_4$, Science **367**, 895 (2020).

[32] C. Liu, Y. Wang, H. Li, Y. Wu, Y. Li, J. Li, K. He, Y. Xu, J. Zhang, and Y. Wang, *Robust Axion Insulator and Chern Insulator Phases in a Two-Dimensional Antiferromagnetic Topological Insulator*, Nat. Mater. **19**, 522 (2020).

[33] Y. Peng, and Y. Xu, *Proximity-Induced Majorana Hinge Modes in Antiferromagnetic Topological Insulators*, Phys. Rev. B **99**, 195431 (2019).

[34] Y. J. Hao, P. Liu, Y. Feng, X.-M. Ma, E. F. Schwier, M.Arita, S. Kumar, C. Hu, R. Lu, M. Zeng *et al.*, *Gapless Surface Dirac Cone in Antiferromagnetomagnetic Topological Insulator $MnBi_2Te_4$*, Phys. Rev. X **9**, 041038 (2019).

[35] H. Li, S.-Y. Gao, S.-F. Duan, Y.-F. Xu, K.-J. Zhu, S.-J. Tian, J.-C. Gao, W.-H. Fan, Z.-C. Rao, J.-C. Huang *et al.*, *Dirac Surface States in Intrinsic Magnetic Topological Insulators $EuSn_2As_2$ and $MnBi_{2n}Te_{3n+1}$*, Phys. Rev. X **9**, 041039 (2019).

[36] Y. J. Chen, L. X. Xu, J. H. Li, Y. W. Li, H. Y. Wang, C. F. Zhang, H. Li, Y. Wu, A. J. Liang, C. Chen *et al.*, *Topological Electronic Structure and Its Temperature Evolution in Antiferromagnetic Topological Insulator $MnBi_2Te_4$*, Phys. Rev. X **9**, 041040 (2019).

[37] Z. Huang, M.-H. Du, J. Yan, and W. Wu, *Native Defects in Antiferromagnetic Topological Insulator $MnBi_2Te_4$*, Phys. Rev. Mater. **4**, 121202 (2020).

[38] H. Tan, and B. Yan, *Distinct Magnetic Gaps between Antiferromagnetic and Ferromagnetic Orders Driven by Surface Defects in the Topological Magnet* $MnBi_2Te_4$, Phys. Rev. Lett. **130**, 126702 (2023).

[39] X. Wu, C. Ruan, P. Tang, F. Kang, W. Duan, and J. Li, *Irremovable Mn-Bi Site Mixing in $MnBi_2Te_4$*, Nano Lett. **23**, 5048 (2023).

[40] S. Xu, Z. Shi, M. Yang, J. Wei, H. Xu, H. Feng, N. Cheng, J. Wang, W. Hao, and Y. Du, *Controllable and Continuous Quantum Phase Transitions in Intrinsic Magnetic Topological Insulator*, arXiv:2503.06044.

[41] A. S. Frolov, D. Y. Usachov, A. V. Tarasov, A. V. Fedorov, K. A. Bokai, I. Klimovskikh, V. S. Stolyarov, A. I. Sergeev, A. N. Larvrov, V. A. Golyashov *et al.*, *Magnetic Dirac Semimetal State of $(Mn,Ge)Bi_2Te_4$*, Commun. Phys. **7**, 180 (2024).

[42] M. Neupane, S.-Y. Xu, L. A. Wray, A. Petersen, R. Shankar, N. Alidoust, C. Liu, A. Fedorov, H. Ji, J. M. Allred *et al.*, *Topological Surface States and Dirac Point Tuning in Ternary Topological Insulators*, Phys. Rev. B **85**, 235406 (2012).

[43] A. M. Shikin, N. L. Zaitsev, T. P. Estyunina, D. A. Estyunin, A. G. Rybkin, D. A. Glazkova, I. I. Klimovskikh, A. V. Eryzhenkov, V. A. Golyashov *et al.*, *Phase Transitions, Dirac and Weyl Semimetal States in $Mn_{1-x}Ge_xBi_2Te_4$*, Sci. Rep. **15**, 1741 (2025).

[44] D. S. Lee, T.-H. Kim, C.-H. Park, C. Y. Chung, Y. S. lim, W.-S. Seo, and H.-H. Park, *Crystal Structure, Properties and Nanostructuring of a New Layered Chalcogenide Semiconductor, $Bi_2MnTe_4$*, CrystEngComm **15**, 5532 (2013).

[45] L. E. Shelimova, O. G. Karpinskii, P. P. Konstantinov, E. S. Avilov, M. A. Kretova, and V. S. Zemskov, *Crystal Structures and Thermoelectric Properties of Layered Compounds in the $ATe–Bi_2Te_3(A=Ge,Sn,Pb)$ Systems*, Inorg. Mater. **40**, 451 (2004).

[46] See Supplemental Material for (1) supercell structures of $Mn_{(1-x)}Ge_xBi_2Te_4$ (MGBT) balancing maintenance of symmetry and energy minimum, (2) band momentum mapping method, evolution of electronic structure and topological properties of antiferromagnetic (AFM) MGBT, (3) detailed information on class-II magnetic Dirac semimetal (MDSM) state, (4) lattice strain application and the feasibility of strain engineering in MGBT, (5) electronic structure and topological properties of ferromagnetic (FM) MGBT, (6) detailed information on class-II Weyl semimetal (WSM) state, (7) detailed information on WSM state with two band inversions (WSM2) and corresponding AFM phase, which also includes Refs. [28,41,42,47-50].

[47] T. P. Estyunina, A. M. Shikin and D. A. Estyunin, A. V. Eryzhenkov, I. I. Klimovskikh, K. A. Bokai, V. A. Golyashov, K. A. Kokh, O. E. Tereshchenko, S. Kumar et al., *Evolution of $Mn_{1−x}Ge_xBi_2Te_4$ Electronic Structure under Variation of Ge Content*. Nanomaterials **13**, 2151 (2023).

[48] W. T. Guo, N. Yang and Z. Huang, and J.-M. Zhang, *Novel Magnetic Topological Insulator $FeBi_2Te_4$ with Controllable Topological Quantum Phase*, J. Mater. Chem. C **11**, 12307 (2023).

[49] W. Ku, T. Berlijn, and C. C. Lee, *Unfolding First-Principles Band Structures*, Phys. Rev. Lett. **104**, 216401 (2010).

[50] Z. Wang, H. Weng, Q. Wu, X. Dai, and Z. Fang, *Three-Dimensional Dirac Semimetal and Quantum Transport in $Cd_3As_2$*, Phys. Rev. B **88**, 125427 (2013).

[51] B. J. Yang, and N. Nagaosa, *Classification of Stable Three-Dimensional Dirac Semimetals with Nontrivial Topology*, Nat. Commun. **5**, 4898 (2014).

[52] S. M. Young, S. Zaheer, J. C. Y. Teo, C. L. Kane, E. J. Mele, and A.M. Rappe, *Dirac Semimetal in Three Dimensions*, Phys. Rev. Lett. **108**, 140405 (2012).

[53] P. Tang, Q. Zhou, G. Xu, and S.-C. Zhang, *Dirac Fermions in an Antiferromagnetic Semimetal*, Nat. Phys. **12**, 1100 (2016).

[54] R. Noguchi, M. Kobayashi, Z. Jiang, K. Kuroda, T. Takahashi, Z. Xu, D. Lee, M. Hirayama, M. Ochi, T. Shirasawa et al., *Evidence for a Higher-Order Topological Insulator in a Three-Dimensional Material Built from van der Waals Stacking of Bismuth-Halide Chains*, Nat. Mater. **20**, 473 (2021).

[55] B. Xie, H.-X. Wang, X. Zhang, P. Zhan, J.-H. Jiang, M. Lu, and Y. Chen, *Higher-Order Band Topology*, Nat. Rev. Phys. **3**, 520 (2021).

[56] X.-J. Luo, X.-H. Pan, C.-X. Liu, and X. Liu, *Higher-Order Topological Phases Emerging from Su-Schrieffer-Heeger Stacking*, Phys. Rev. B **107**, 045118 (2023).

[57] W. A. Benalcazar, B. A. Bernevig. and T. L. Hughes, *Electric Multipole Moments, Topological Multipole Moment Pumping, and Chiral Hinge States in Crystalline Insulators*, Phys. Rev. B **96**, 245115 (2017).

[58] P. E. Blöchl, *Projector Augmented-Wave Method*, Phys. Rev. B **50**, 17953 (1994).

[59] G. Kresse, and J. Furthmüller, *Efficient Iterative Schemes for ab initio Total-Energy Calculations Using a Plane-Wave Basis set*, Phys. Rev. B **54**, 11169 (1996).

[60] J. P. Perdew, K. Burke, and Ernzerhof, M. *Generalized Gradient Approximation Made Simple*, Phys. Rev. Lett. **77**, 3865 (1996).

[61] S. Grimme, J. Antony, S. Ehrlich, and H. Krieg, *A Consistent and Accurate ab initio Parametrization of Density Functional Dispersion Correction (DFT-D) for the 94 Elements H-Pu*, Chem. Phys. **132**, 154104 (2010).

[62] M. J. Han, T. Ozaki, and J. Yu, *O(N) LDA+U Electronic Structure Calculation Method Based on the Nonorthogonal Pseudoatomic Orbital Basis*, Phys. Rev. B **73**, 045110 (2006).

[63] I. Souza, N. Marzari, and D. Vanderbilt, *Maximally localized Wannier functions for entangled energy bands*, Phys. Rev. B **65**, 035109 (2001).

[64] N. Marzari, A. A. Mostofi, J. R. Yates, I. Souza, and D. Vanderbilt, *Maximally Localized Wannier Functions: Theory and Applications*, Rev. Mod. Phys. **84**, 1419 (2012).

[65] Q. Wu, S. Zhang, H. F. Song, M. Troyer, and A. Soluyanov, *WannierTools: An Open-Source Software Package for Novel Topological Material,*. Comput. Phys. Commun. **224**, 405 (2018).

[66] G.-X. Zhi, C. Xu, S.-Q. Wu, F. Ning, and C. Cao, *WannSymm: A Symmetry Analysis Code for Wannier Orbitals*, Comput. Phys. Commun. **271**, 108196 (2022).